\documentclass[12pt,draftclsnofoot,onecolumn]{IEEEtran}
\usepackage{comment,url}
\usepackage[dvips]{graphicx}
\usepackage{caption,subfigure,color}
\usepackage{float}
\usepackage{setspace}
\usepackage{amsmath}
\usepackage{amssymb}
\usepackage{algorithm}
\usepackage{algorithmic}
\usepackage{cite}
\usepackage{array}

\newcommand{\mysize}{7cm}
\allowdisplaybreaks[4]
\long\def\symbolfootnote[#1]#2{\begingroup%
\def\thefootnote{\fnsymbol{footnote}}\footnote[#1]{#2}\endgroup}

\begin{document}
%
\title{The Distance Distribution between Mobile Node and Reference Node in Regular Hexagon\\
\thanks{Corresponding author: Fei Tong. This work is supported in part by the National Natural Science Foundation of China under Grants 61971131 and 61702452, in part by ``Zhishan'' Scholars Programs of Southeast University, in part by the Ministry of Educations Key Lab for Computer Network and Information Integration, Southeast University, Nanjing, China, and in part by the Fundamental Research Funds for the Central Universities.
}
}

\author{\IEEEauthorblockN{Ziyan Zhu\IEEEauthorrefmark{1} and Fei Tong\IEEEauthorrefmark{1}\IEEEauthorrefmark{2}
}\\
\IEEEauthorblockA{\IEEEauthorrefmark{1}Southeast University, Nanjing, Jiangsu, China}\\
\IEEEauthorblockA{\IEEEauthorrefmark{2}Purple Mountain Laboratories, Nanjing, Jiangsu, China}
}

\maketitle
\begin{abstract}
This paper presents a new method to obtain the distance distribution between the mobile node and any reference node in a regular hexagon. The existing distance distribution research mainly focuses on static network deployment and ignores node mobility. This paper studies the distribution of node distances between mobile node and any reference node. A random waypoint (RWP) migration model is adopted for mobile node. The Cumulative Distribution Function (CDF) of the distance between any reference node (inside or outside the regular hexagon) and  the mobile node (inside the regular hexagon) is derived. The validity of the results is verified by simulation.
\end{abstract}


\section{System Model}

\begin{figure}[htbp]
	\centering
	\includegraphics{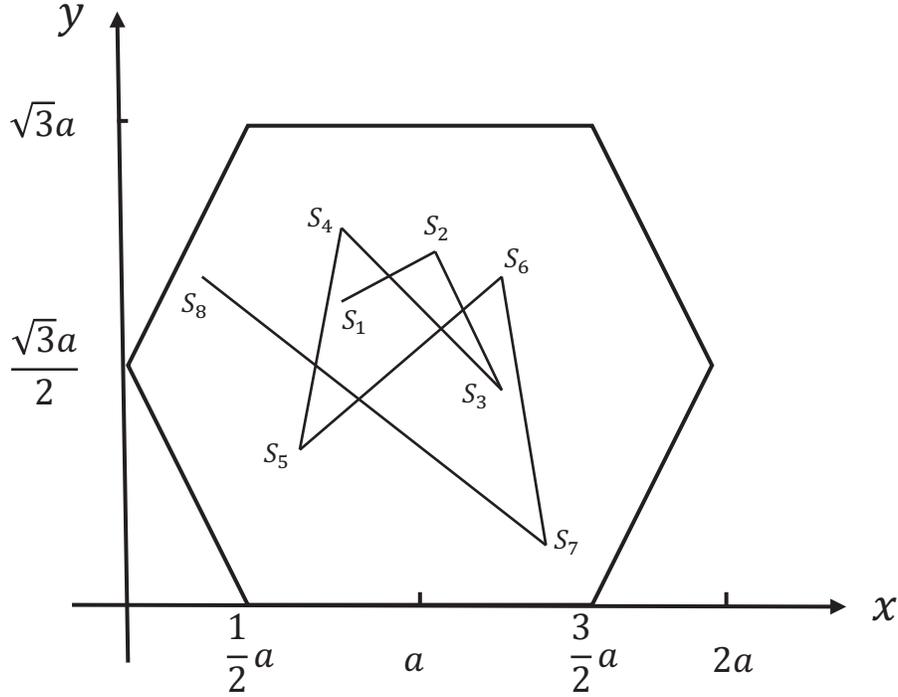}
	\caption{RWP Node Trajectory}
	\label{rwp}
\end{figure}

In mobility management, RWP is a random model that simulates the movement of mobile users and how their position, speed, and acceleration change over time. Due to its simplicity and wide availability, it is one of the most popular mobile models for evaluating mobile ad hoc network performance. Figure~\ref{rwp} shows an example to illustrate the RWP trajectory of the node in a regular hexagon area.

Considering the Cartesian coordinate system established in Fig.~\ref{rwp}, below is to show the complete process of RWP in the regular hexagon area: given the simulation time $T$, randomly and uniformly generate the starting point $S_{1}$ of the mobile node. Then the mobile node will select a destination point $D_{1}$ and randomly and uniformly select a speed $\nu$ from $[\nu_{min},\nu_{max}]$. The moving time between the destination point $D_{1}$ and the starting point $S_{1}$ is $ \frac{D_{1}-S_{1}}{\nu}$, which completes the $1$-period. To simplify the model, we do not consider the effect of pause time $t_{p}$ and let $t_{p} = 0$. If there is remaining simulation time, the mobile node immediately selects the next destination point $D_{2}$, and randomly and uniformly select a speed from $[\nu_{min},\nu_{max}]$, then move from $S_{2}$ to $D_{2}$. At this time, $D_{1}$ is the starting point $S_{2}$ in the $2$-period. The mobile node has a starting point $S_{k}$ and a destination point $D_{k}$ in each $k$-period movement.\par

\section{Distance Distribution from An Arbitrary Reference Node to A Mobile Node}
\begin{figure}[htbp]
	\vspace{-1em}
	\centering
	\includegraphics[width=0.7\textwidth]{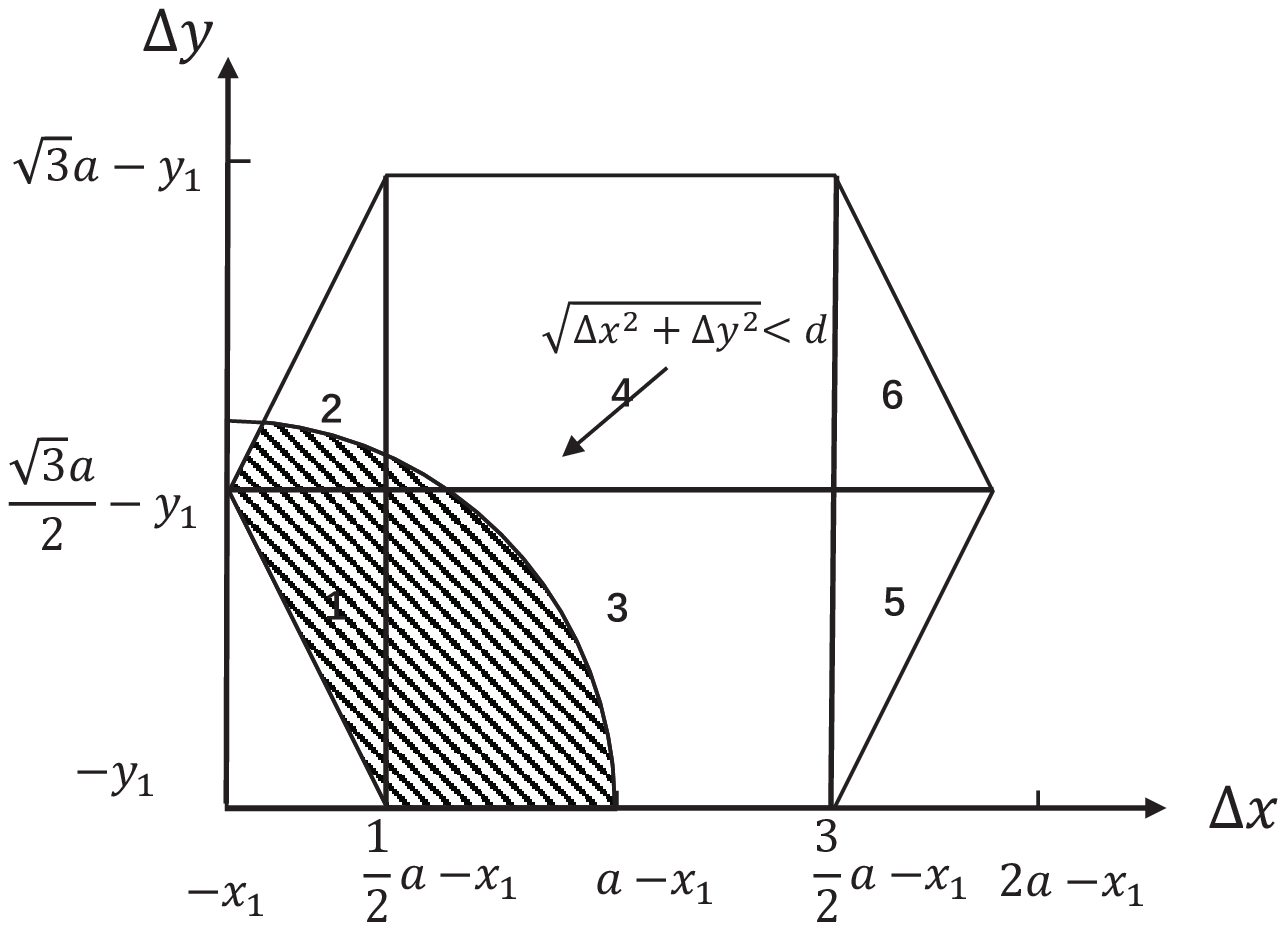}
	\vspace{-1.5em}
	\caption{Integration area.}
	\label{rwp-area}
\end{figure}

As shown in Fig.~\ref{rwp}, we consider a regular hexagonal cell in a cellular network. In this paper we investigate the distribution of the distance between an arbitrary reference node denoted as $N_{1} = (x_{1},\;y_{1})$ and a mobile node denoted as $N = (x,\; y)$. The random variable of distance between the moblie node and the reference node is defined as
\begin{equation}
	D_{m} = \sqrt{\Delta x^{2}+\Delta y^{2}}~,
	\label{eq1}
\end{equation}
where $\Delta x = x-x_{1}$, and $\Delta y = y-y_{1}$. Let $\Delta X$ and $\Delta Y$ denote the random variables of distance between a moblie node and a reference node in $x$-axis and $y$-axis, respectively. The Cumulative Distribution Function (CDF) of $D_{m}$ denoted as $F_{D_{m}}(d)$, can be represented as
\begin{equation}
	\label{eq2}
	\begin{split}
		F_{D_{m}}(d) &= {\rm Pr}(D_{m}<d)\\&= \frac{\iint_{\omega}f_{\Delta X}(\Delta x)f_{\Delta Y}(\Delta y)\mathrm{d}\Delta x\mathrm{d}\Delta y}{\iint_{O}f_{\Delta X}(\Delta x)f_{\Delta Y}(\Delta y)\mathrm{d}\Delta x\mathrm{d}\Delta y}~,
	\end{split}
\end{equation}
where $O$ denotes the regular hexagon area, and $\omega$ denotes the integration area between the circle area defined by $\sqrt{\Delta x^{2}+\Delta y^{2}} < d$ and $O$, i.e., the shaded part in Fig.~\ref{rwp-area}. $f_{\Delta X}(\Delta x)$ and $f_{\Delta Y}(\Delta y)$ denote the PDFs of $\Delta X$ and $\Delta Y$, respectively. Let $X$ and $Y$ denote the random variables of moblie node coordinates in $x$-axis and $y$-axis, respectively. $f_{X}(x)$ and $f_{Y}(y)$ denote the PDFs of $X$ and $Y$, respectively. Since $f_{\Delta X}(\Delta x)$ and $f_{\Delta Y}(\Delta y)$ are related to $f_{X}(x)$ and $f_{Y}(y)$, we below derive $f_{\Delta X}(\Delta x)$, $f_{\Delta Y}(\Delta y)$ by $f_{X}(x)$ and $f_{Y}(y)$, respectively. Let the side length of the regular hexagon be $a$, where $X\in \left [ 0, 2a \right]$ and $Y\in \left [ 0,\sqrt{3}a \right]$. 

\subsection{$f_{X}(x)$}
Let the random variables $S$ and $D$ denote the starting and destination coordinates of a movement period in $x$-axis, respectively. The CDFs $F_{S}(s)$ and $F_{D}(d)$ in $x$-axis can be obtained by the area ratio method. $S$ and $D$ have the same distribution. Therefore,
\begin{equation}
	f_{S}(s) =f_{D}(d)= \begin{cases}
		\frac{4s}{3a^{2}}& s\in [0,\frac{1}{2}a] \\
		\frac{2}{3a}& s\in(\frac{1}{2}a,\frac{3}{2}a]  \\
		\frac{4(2a-s)}{3a^{2}} & s\in(\frac{3}{2}a,2a]
	\end{cases}~.
	\label{eq4}
\end{equation}

In order to derive $f_{X}(x)$, we first calculate the $F_{X}(x) = P(X\leq x)$, which denotes the probability that the mobile node is located within $\left [ 0, x \right ]$ at an arbitrary instant of time in $x$-axis. According to\cite{Bettstetter2003257}, we have $P(X\leq x) = \frac{E[L_{x}]}{E_{L}}$, where $L$ is the movement distance in each period, $L_{x}$ is the line segment of $L$ whose coordinate is less than $x$, $E[L]$ is the mathematical expectation of $L$ in each period, and $E[L_{x}]$ is the mathematical expectation of $L_{x}$.

\subsubsection{$E(L)$}
let $l(s,d)$ denote the value of the random variable $L$ if $S = s$ and $D = d$. Because of the symmetry of S and D, we restrict the calculation to periods with $s < d$ and then multiply the result by a factor of 2. We have:
\begin{equation}
	E(L)=2\int_{s=0}^{2a}\int_{d=s}^{2a}l(s,d)f_{S}(s)f_{D}(d)\mathrm{d}d \mathrm{d}s
	\label{eq5}
\end{equation}
Finally, we derive $E(L) = \frac{71}{135}a$. The detail of the derivation can be found in Appendix \ref{sec:APPENDIX-A}.

\subsubsection{$E(L_{x})$}
let $l_{x}(s,d)$ denotes the value of the random variable $L_{x}$ if $S = s$ and $D = d$. $l_{x}(s,d)$ represents the line segment from $s$ to $d$ whose coordinate is less than $x$. Similarly, $s < d$ and multiply the result by a factor of 2. We have:
\begin{equation}
	E(L_{x})=2\int_{s=0}^{2a}\int_{d=s}^{2a}l_{x}(s,d)f_{S}(s)f_{D}(d)\mathrm{d}d \mathrm{d}s~.
	\label{eq7}
\end{equation}
By calculating and simplifying~\eqref{eq7}, we derive $\frac{1}{2}E(L_{x})=$
\begin{equation}
	\begin{cases}
		\frac{2x^{3}}{9a^{2}}-\frac{4x^{5}}{45a^{4}}
		& x\in [0,\frac{1}{2}a] \\
		\frac{4a}{135}-\frac{7x}{36}+\frac{4x^{2}}{9a}-\frac{4x^{3}}{27a^{2}}  & x\in (\frac{1}{2}a,\frac{3}{2}a] \\
		\frac{359a^{5}-1200a^{4}x+1560a^{3}x^{2}}{270a^{4}}+\\ \frac{-900a^{2}x^{3}+240ax^{4}-24x^{5}}{270a^{4}} & x\in (\frac{3}{2}a,2a] \\
	\end{cases}~.
\end{equation}
The detail of the derivation can be found in Appendix \ref{APPENDIX-B}.

\subsubsection{$f_{X}(x)$}
According to $F_{X}(x) = P(X\leq x) = \frac{E[L_{x}]}{E_{L}}$,the PDF $f_{X}(x)$ of $X$ is
\begin{equation}
	\label{pdfx2}
	f_{X}(x)=\begin{cases}
		\frac{12(15a^{2}x^{2}-10x^{4})}{71a^{5}}
		& x\in [0,\frac{1}{2}a] \\
		\frac{-105a^{2}+480ax-240x^{2}}{142a^{3}}  & x\in (\frac{1}{2}a,\frac{3}{2}a] \\
		\frac{-1200a^{4}+3120a^{3}x}{71a^{5}}+\\ \frac{-2700a^{2}x^{2}+960ax^{3}-120x^{4}}{71a^{5}} & x\in (\frac{3}{2}a,2a] \\
	\end{cases}~.
\end{equation}
\subsection{$f_{Y}(y)$}
Let the random variables $S$ and $D$ denote the starting and destination points of a movement period in $y$-axis. We can derive $f_{S}(s)$ as follows:
\begin{equation}
	f_{S}(s) = \begin{cases}
		\frac{2(\sqrt{3}a+2s)}{9a^{2}} & s\in [0,\frac{\sqrt{3}}{2}a] \\
		\frac{6\sqrt{3}a-4s}{9a^{2}} & s\in [\frac{\sqrt{3}}{2}a,\sqrt{3}a]
	\end{cases}~.
	\label{eqy}
\end{equation}
Similarly, $f_{S}(s) = f_{D}(d)$.
In order to derive $f_{Y}(y)$, we first calculate the $F_{Y}(y) = P(Y\leq y)$, which denotes the probability that the mobile node is located within $\left [ 0, y \right ]$ at an arbitrary instant of time. Similarly, we have $P(Y\leq y) = \frac{E[L_{y}]}{E_{L}}$, where $L$ is the movement distance in each period in $y$-axis, $L_{y}$ is the line segment of $L$ whose coordinate is less than $y$, $E[L]$ is the mathematical expectation of $L$ in $y$-axis in each period, and $E[L_{y}]$ is the mathematical expectation of $L_{y}$.

\subsubsection{$E(L)$}
let $l(s,d)$ denote the value of the random variable $L$ if $S = s$ and $D = d$. We have:
\begin{equation}
	E(L)=2\int_{s=0}^{\sqrt{3}a}\int_{d=s}^{\sqrt{3}a}l(s,d)f_{S}(s)f_{D}(d)\mathrm{d}d \mathrm{d}s~.
	\label{eq3}
\end{equation}
According to \eqref{eq3}, we derive $E(L) = \frac{41a}{45\sqrt{3}}$. The derivation detail can be found in Appendix \ref{sec:APPENDIX-A}.

\subsubsection{$E(L_{y})$}
let $l_{y}(s,d)$ denote the value of the random variable $L_{y}$ if $S = s$ and $D = d$. We have:
\begin{equation}
	E(L_{y})=2\int_{s=0}^{\sqrt{3}a}\int_{d=s}^{\sqrt{3}a}l_{y}(s,d)f_{S}(s)f_{D}(d)\mathrm{d}d \mathrm{d}s~.
	\label{yely}
\end{equation}
By calculating and simplifying \eqref{yely}, we derive $\frac{1}{2}E(L_{y})=$
\begin{equation}
	\begin{cases}
		\frac{y^{2}(45\sqrt{3}a^{3}+10a^{2}y-10\sqrt{3}ay^{2}-4y^{3})}{405a^{4}} & y\in [0,\frac{\sqrt{3}}{2}a] \\
		\frac{45\sqrt{3}a^{5}-360a^{4}y+450\sqrt{3}a^{3}y^{2}}{810a^{4}}+\\ \frac{-460a^{2}y^{3}+60^{3}ay^{4}-8y^{5}}{810a^{4}} & y\in (\frac{\sqrt{3}}{2}a,\sqrt{3}a]
	\end{cases}~.
\end{equation}
The derivation detail can be found in Appendix \ref{APPENDIX-B}.

\subsubsection{$f_{Y}(y)$}
According to $F_{Y}(y) = P(Y\leq y) = \frac{E[L_{y}]}{E_{L}}$. The PDF $f_{Y}(y)$ of $Y$ is
\begin{equation}
	\label{pdfy2}
	f_{Y}(y)=\begin{cases}
		\frac{20y(27a^{3}+3\sqrt{3}a^{2}y-12ay^{2}-2\sqrt{3}y^{3})}{369a^{5}} & y\in [0,\frac{\sqrt{3}}{2}a]\\
		\frac{-360a^{4}+900\sqrt{3}a^{3}y-138a^{2}y^{2}}{123\sqrt{3}a^{5}}+\\
		\frac{240\sqrt{3}ay^{3}-40y^{4}}{123\sqrt{3}a^{5}} & y\in (\frac{\sqrt{3}}{2}a,\sqrt{3}a]
	\end{cases}~.
\end{equation}

\subsection{Distance Distribution from An Arbitrary Reference Node to A Mobile Node}\label{mobile-cdf}
According to \eqref{eq2}, we need to know $f_{\Delta X}(\Delta x)$ and $f_{\Delta Y}(\Delta y)$ to get $F_{D_{m}}(d)$. According to \eqref{eq1}, we have:
\begin{equation}
	\left\{\begin{split}
		\Delta x &= x-x_{1} \quad x\in [0,2a],\Delta x\in [-x_{1},2a-x_{1}] \\
		\Delta y &= y-y_{1} \quad y\in [0,\sqrt{3}a],\Delta y\in [-y_{1},\sqrt{3}a - y_{1}]
	\end{split}\right..
\end{equation}
Substitute variable $x$ to $\Delta x$:
\begin{equation}
	\begin{split}
		F_{\Delta X}(\Delta x) &= {\rm Pr}(\Delta X\leq \Delta x)={\rm Pr}(x-x_{1}\leq \Delta x)\\
		&={\rm Pr}(x\leq \Delta x+x_{1})=F_{X}(\Delta x+x_{1})~.
	\end{split}
\end{equation}
Because of $f_{\Delta X}(\Delta x) = {F_{\Delta X}}'(\Delta x)$, it is easy to derive:
\begin{equation}
	\begin{split}
		f_{\Delta X}(\Delta x) = {F_{X}}'(\Delta x+x_{1})=f_{X}(\Delta x+x_{1})~.
	\end{split}
\end{equation}
Let $\Delta x$ value range be denoted by:
\begin{equation}
	\left\{\begin{split}
		\phi_{1} &= \{\Delta x | \Delta x\in [-x_{1},\frac{1}{2}a-x_{1}]\}\\
		\phi_{2} &= \{\Delta x | \Delta x\in (\frac{1}{2}a-x_{1},\frac{3}{2}a-x_{1}]\}\\
		\phi_{3} &= \{\Delta x | \Delta x\in (\frac{3}{2}a-x_{1},2a-x_{1}]\}
	\end{split}\right..
\end{equation}
Then, according to \eqref{pdfx2}, we derive $f_{\Delta X}(\Delta x)=$
\begin{equation}
	\begin{split}
		\begin{cases}
			\frac{12(15a^{2}(\Delta x+x_{1})^{2}-10(\Delta x+x_{1})^{4})}{71a^{5}}
			&\Delta x\in \phi_{1} \\
			\frac{-105a^{2}+480a(\Delta x+x_{1})-240(\Delta x+x_{1})^{2}}{142a^{3}}  &\Delta x\in \phi_{2} \\
			\frac{-1200a^{4}+3120a^{3}(\Delta x+x_{1})-2700a^{2}(\Delta x+x_{1})^{2}}{71a^{5}}+\\\frac{960a(\Delta x+x_{1})^{3}-120(\Delta x+x_{1})^{4}}{71a^{5}} &\Delta x\in \phi_{3} \\
		\end{cases}~.
	\end{split}
\end{equation}
Let $\Delta y$ value range be denoted by:
\begin{equation}
	\left\{
	\begin{split}
		\psi_{1} &= \{\Delta y | \Delta y\in [-y_{1},\frac{\sqrt{3}}{2}a-y_{1}]\}\\
		\psi_{2} &= \{\Delta y | \Delta y\in (\frac{\sqrt{3}}{2}a-y_{1},\sqrt{3}a-y_{1}]\}
	\end{split}\right..
\end{equation}
Similarly, according to \eqref{pdfy2}, $f_{\Delta Y}(\Delta y)$ can be derived by variable substitution. So $f_{\Delta Y}(\Delta y)=$
\begin{equation}
	\begin{split}
		\begin{cases}
			\frac{540a^{3}(\Delta y+y_{1})+60\sqrt{3}a^{2}(\Delta y+y_{1})^{2}}{369a^{5}}+\\\frac{-240a(\Delta y+y_{1})^{3}-40\sqrt{3}(\Delta y+y_{1})^{4}}{369a^{5}} & \Delta y\in \psi_{1}\\
			\frac{-360a^{4}+900\sqrt{3}a^{3}(\Delta y+y_{1})-138a^{2}(\Delta y+y_{1})^{2}}{123\sqrt{3}a^{5}}+\\
			\frac{240\sqrt{3}a(\Delta y+y_{1})^{3}-40(\Delta y+y_{1})^{4}}{123\sqrt{3}a^{5}} & \Delta y\in \psi_{2}
		\end{cases}~.
	\end{split}
\end{equation}
Then, $F_{D_{m}}(d)$ in \eqref{eq2} can be derived by zoning law over the integral region.

\section{Simulation Result}\label{subsec:result}
In this section, we verify the distance distribution between mobile node and reference node derived in the above sections by MATLAB simulation platform, which allows us to simulate the movement characteristics of the mobile node and record its locations. We compare the simulation results with the theoretical results under different parameters.

The mobile node is modeled with RWP to move in a regular hexagonal with side length of one, and the pause time $t_{p} = 0$. The position of mobile node at each moment is recorded in the time interval of one second. In each time interval, we obtain the distances between the mobile node and the reference node. After the simulation period $t$ is over, the obtained distance samples are applied to estimate the CDF. We use the MATLAB function ``ecdf()'' to calculate the CDF.

\begin{figure}[htbp]
	\centering
	\includegraphics[width=9.5cm]{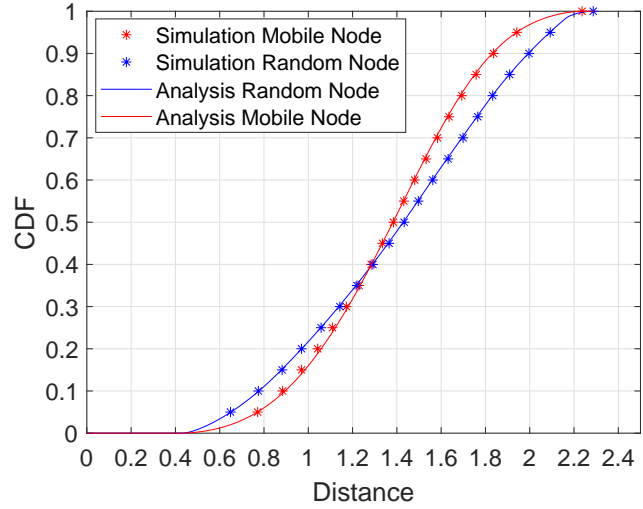}
	\caption{Comparison of distance distribution from origin of coordinates to random node and mobile node}
	\label{rand_move}
\end{figure}

Figure.~\ref{rand_move} compare of distance distribution between random node and mobile node in regular hexagon with side length of one. The velocity $v$ of mobile node is randomly selected from the interval $[0.01 s^{-1}, 0.05 s^{-1}]$, and the duration of simulation period is set to $t = 100000$ s. As can be seen from the figure, the distance distribution from origin of coordinates to random node and mobile node is greatly different. Therefore, it is necessary to study the mobility of nodes.

\begin{figure}[H]
	\begin{minipage}[t]{0.5\textwidth}
		\centering
		\includegraphics[scale=0.5]{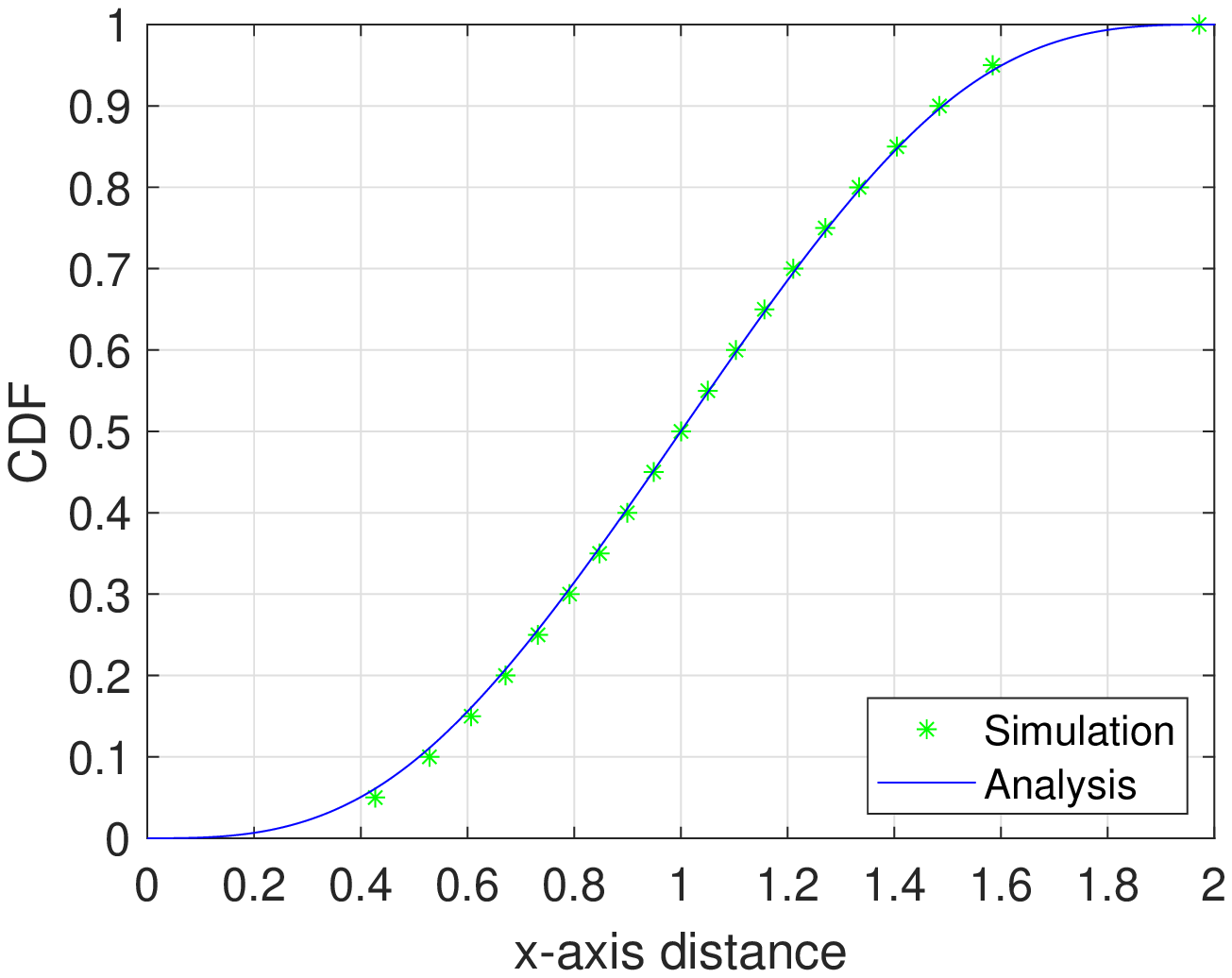}
		\caption{The analysis and simulation of $F_{X}(x)$\label{fig:1}}
	\end{minipage}
	\qquad
	\begin{minipage}[t]{0.5\textwidth}
		\centering
		\includegraphics[scale=0.5]{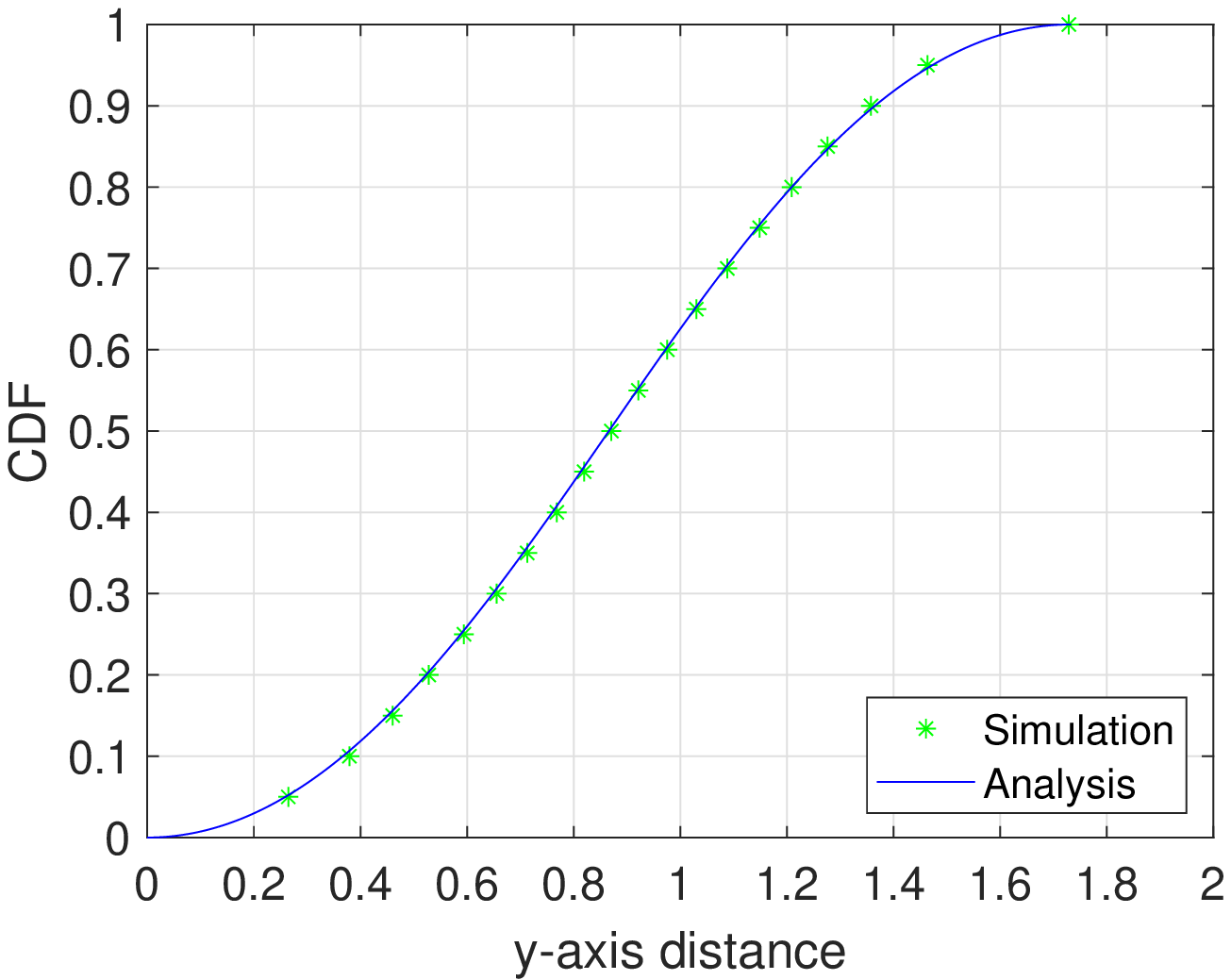}
		\caption{The analysis and simulation of $F_{Y}(y)$ \label{fig:2}}
	\end{minipage}
\end{figure}

\begin{figure}[H]
	\begin{minipage}[t]{0.5\textwidth}
		\centering
		\includegraphics[scale=0.5]{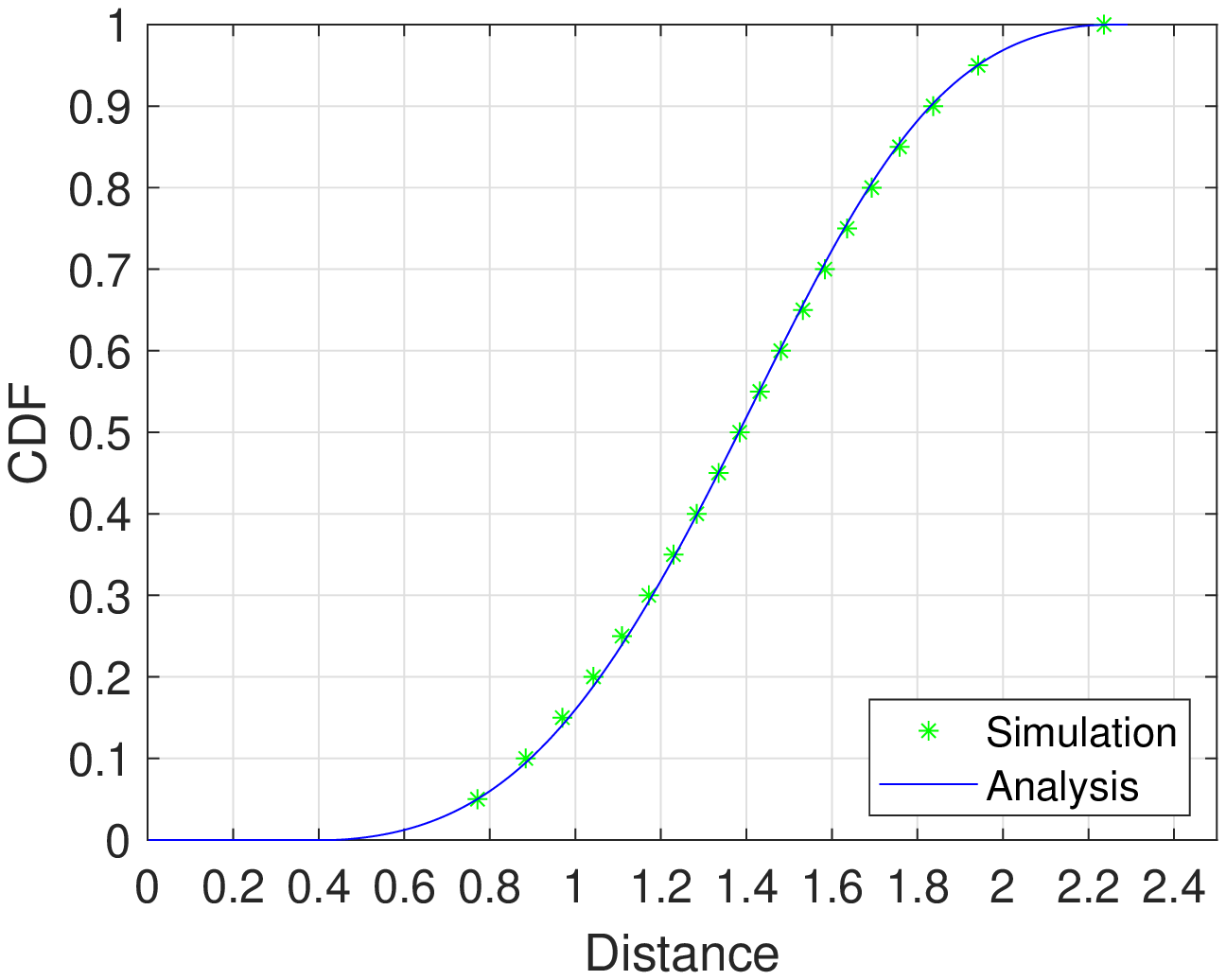}
		\caption{The CDF of the distance from reference node (0,0) to the mobile node.
		\label{fig:3}}
	\end{minipage}
	\qquad
	\begin{minipage}[t]{0.5\textwidth}
		\centering
		\includegraphics[scale=0.5]{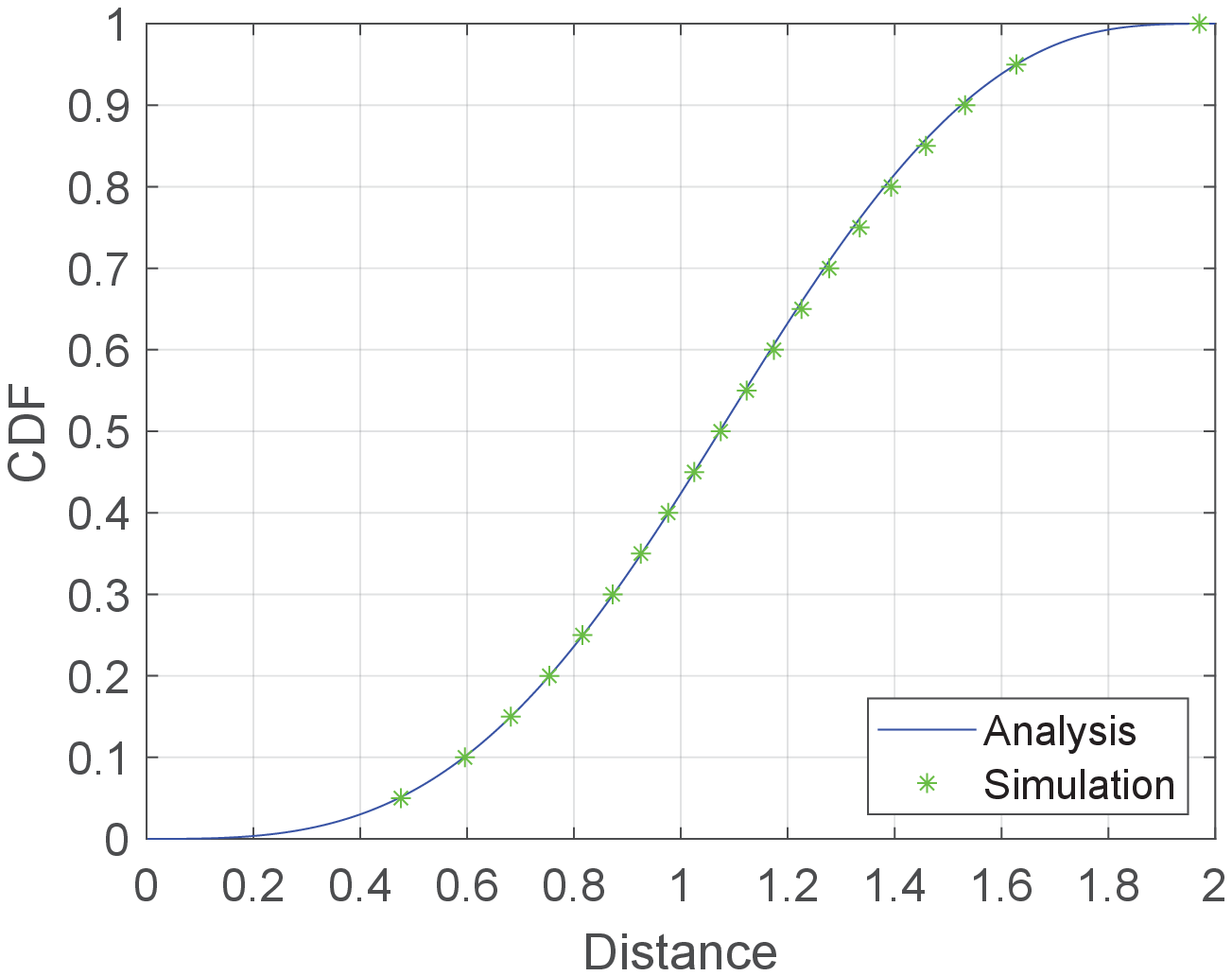}
		\caption{The CDF of the distance from reference node ($\frac{1}{2}$,0) to the mobile node.
		\label
			{fig:4}}
	\end{minipage}
\end{figure}

\begin{figure}[H]
	\begin{minipage}[t]{0.5\textwidth}
		\centering
		\includegraphics[scale=0.5]{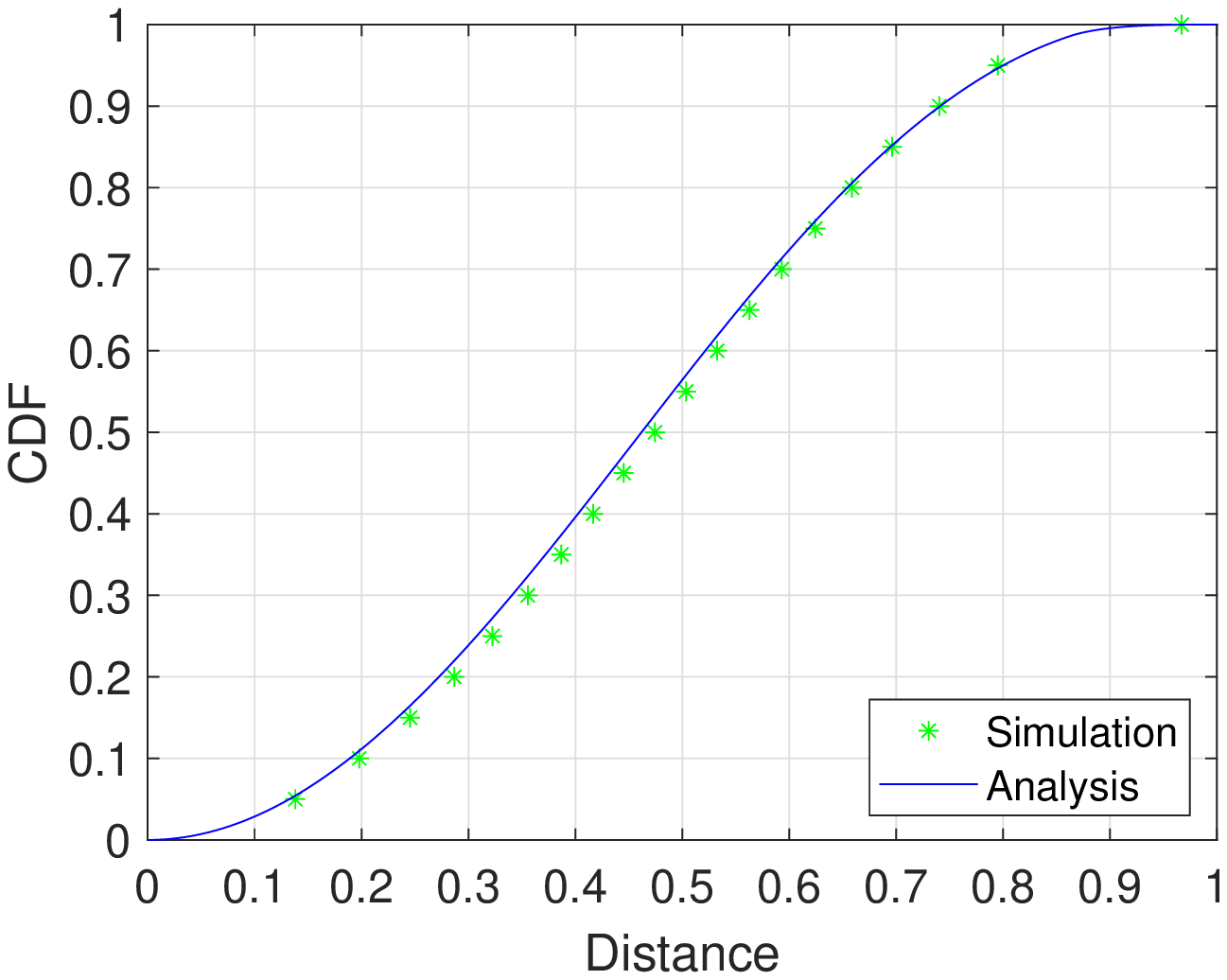}
		\caption{The CDF of the distance from reference node (1,$\frac{\sqrt{3}}{2}$) to the mobile node.
		\label{fig:5}}
	\end{minipage}
	\qquad
	\begin{minipage}[t]{0.5\textwidth}
		\centering
		\includegraphics[scale=0.5]{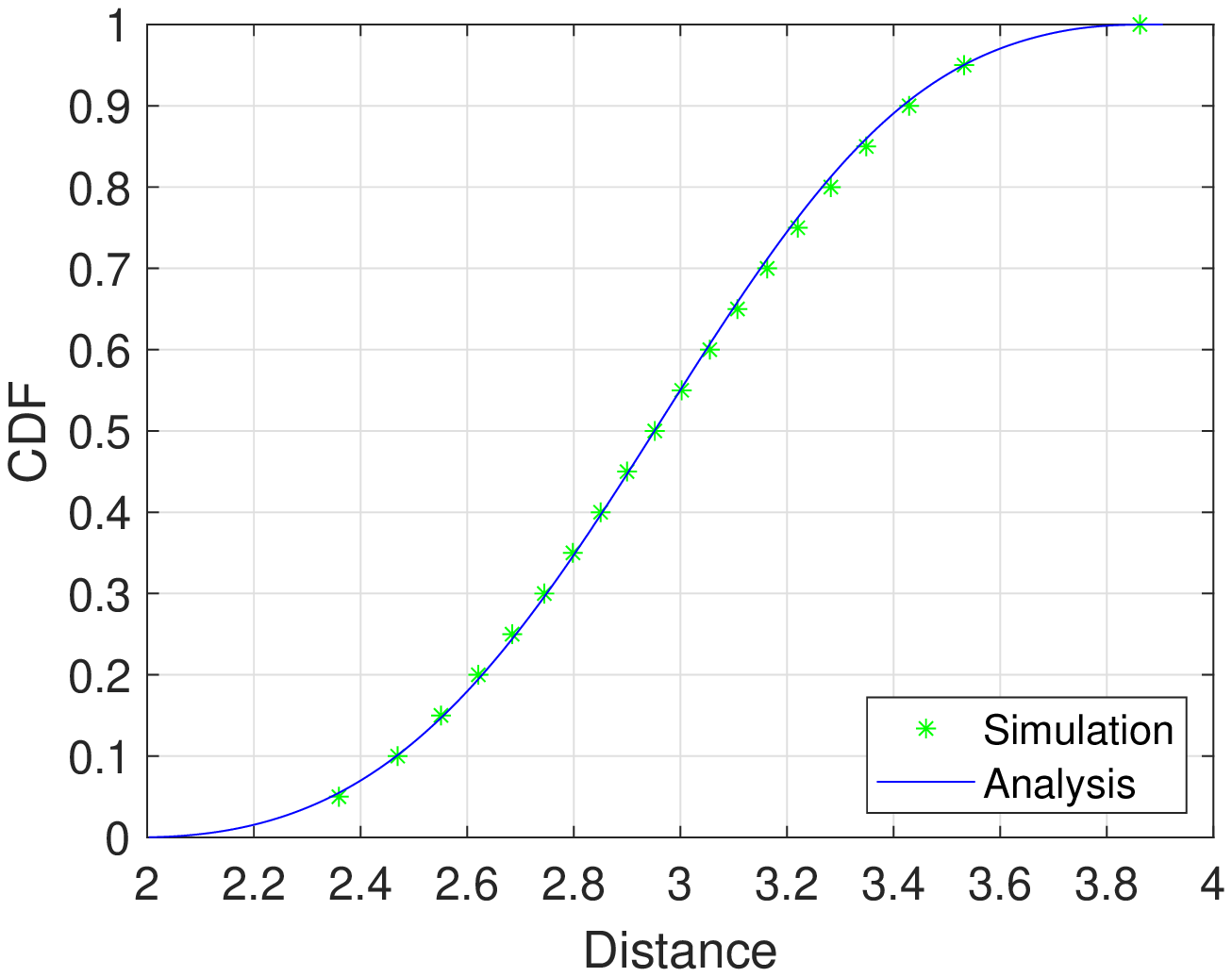}
		\caption{The CDF of the distance from reference node (3,3) to the mobile node.
		\label{fig:6}}
	\end{minipage}
\end{figure}

\begin{figure}[htbp]
	\centering
	\includegraphics[width=9.5cm]{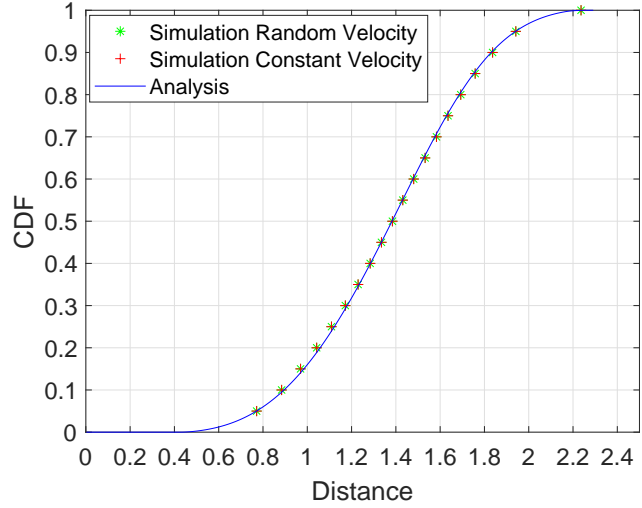}
	\caption{Analytical CDF vs simulation results under different velocity models}
	\label{fig7}
\end{figure}

Figure.~\ref{fig:1} and Fig.~\ref{fig:2} show the obtained $F_{X}(x)$ and $F_{Y}(y)$ with $v$ randomly selected from the interval $[0.01 s^{-1}, 0.05 s^{-1}]$ and $t = 100000$ s. As can be observed,the results of our analysis and simulation are very close. Figure.~\ref{fig:3} - \ref{fig:6} show the obtained results with different reference nodes with $v$ randomly selected from the interval $[0.01 s^{-1}, 0.05 s^{-1}]$ and $t = 100000$ s. As can be seen from the figure, there is a slight deviation between the simulation results and the analysis results. The authors in \cite{Bettstetter2003257}, pointed out that the two-dimensional movement is composed of two dependent one-dimensional movements. The speed of a node projected along the $x$-axis is not constant and it is different from the speed along the $y$-axis. Therefore, an approximation $f_{\Delta X\Delta Y}(\Delta x,\Delta y)\approx  f_{\Delta X}(\Delta x)f_{\Delta Y}(\Delta y)$ is adopted within the acceptable range of differences. Figure.~\ref{fig7} compares of Analytical CDF vs simulation results with random velocity with $v$ randomly selected from the interval $[0.01 s^{-1}, 0.05 s^{-1}]$ and constant velocity with $v = 0.03 s^{-1}$.  It shown that our analysis is available for both constant and random speeds, and different speeds model have little influence on the results.

\section{Conclusions}
In physical layer security, the distance between communication nodes is a very important parameter, which will directly affect physical layer security performance. Considering the importance of distance distribution, we present a new method to obtain the distance distribution between the mobile node and any reference node in a regular hexagon. The distance distribution between mobile node and reference node proposed in this paper can be applied to analyze the physical layer security performance of base station and mobile user, and apply it to physical layer security performance analysis in the future.
\appendix
\subsection{APPENDIX-A}\label{sec:APPENDIX-A}
\subsubsection{$E(L)$ in $x$-axis}
Let $\theta_{1}=\{s | s\in [0,\frac{1}{2}a]\}$, $\theta_{2}=\{s | s\in (\frac{1}{2}a,\frac{3}{2}a]\}$, and $\theta_{3}=\{s | s\in (\frac{3}{2}a,2a]\}$. Since $f_{S}(s)$ and $f_{D}(d)$ are piecewise functions, we have:
\begin{equation}
	\label{el}
	E(L) =
	\begin{cases}
		2\int_{0}^{\frac{1}{2}a}\int_{s}^{\frac{1}{2}a}(d-s)f_{S_{1}}(s)f_{D_{1}}(d)\mathrm{d}d\mathrm{d}s &  s,d\in \theta_{1} \\
		2\int_{0}^{\frac{1}{2}a}\int_{\frac{1}{2}a}^{\frac{3}{2}a}(d-s)f_{S_{1}}(s)f_{D_{2}}(d)\mathrm{d}d\mathrm{d}s & d \in \theta_{2}, s \in \theta_{1} \\
		2\int_{\frac{1}{2}a}^{\frac{3}{2}a}\int_{s}^{\frac{3}{2}a}(d-s)f_{S_{2}}(s)f_{D_{2}}(d)\mathrm{d}d\mathrm{d}s & s,d\in\theta_{2} \\
		2\int_{0}^{\frac{1}{2}a}\int_{\frac{3}{2}a}^{2a}(d-s)f_{S_{1}}(s)f_{D_{3}}(d)\mathrm{d}d\mathrm{d}s & d \in\theta_{3}, s \in \theta_{1} \\
		2\int_{\frac{1}{2}a}^{\frac{3}{2}a}\int_{\frac{3}{2}a}^{2a}(d-s)f_{S_{2}}(s)f_{D_{3}}(d)\mathrm{d}d\mathrm{d}s & d \in\theta_{3}, s\in\theta_{2} \\
		2\int_{\frac{3}{2}a}^{2a}\int_{s}^{2a}(d-s)f_{S_{3}}(s)f_{D_{3}}(d)\mathrm{d}d\mathrm{d}s & s,d \in\theta_{3}
	\end{cases}~,
\end{equation}
where according to \eqref{eq4}, $f_{S_{1}}(s) = \frac{4s}{3a^{2}}$, $ f_{S_{2}}(s) = \frac{2}{3a}$, and $ f_{S_{3}}(s) = \frac{4(2a-s)}{3a^{2}}$. $f_{D_{1}}(d)$, $f_{D_{2}}(d)$, and $f_{D_{3}}(d)$ are the same as $f_{S_{1}}(s)$, $f_{S_{2}}(s)$, and $f_{S_{3}}(s)$, respectively. According to \eqref{eq5}, we restrict the calculation to the period with $s < d$ and then multiply the result by a factor of 2. Since $f_{S}(s)$ and $f_{D}(d)$ are piecewise functions and $s, d \in {\theta_{1}, \theta_{2}, \theta_{3}}$. Therefore, there are six cases, and $E(L)$ is the sum of the six cases. In the first case $s, d \in \theta_{1}$, the range of integration of variable $s$ is from $0$ to $\frac{1}{2}a$, and because of $s<d$, the range of integration of variable $d$ is from $s$ to $\frac{1}{2}a$. In the second case, $s \in \theta_{1}$ and $d \in \theta_{2}$, the range of integration of variable $s$ is from $0$ to $\frac{1}{2}a$, and the lower limit of integration of variable $d$ is not $s$ but $\frac{1}{2}a$, because $d$ must be bigger than $s$. The other four cases are also analyzed in the same way. Finally, by calculating \eqref{el}, we derive $E(L) = \frac{71}{135}a$.

\subsubsection{$E(L)$ in $y$-axis}
Let $\vartheta_{1}=\{s | s\in [0,\frac{\sqrt{3}}{2}a]\}$, and $\vartheta_{2}=\{s | s\in (\frac{\sqrt{3}}{2}a,\sqrt{3}a]\}$. Then we have:
\begin{equation}
	\label{yel}
	E(L) =
	\begin{cases}
		2\int_{s=0}^{\frac{\sqrt{3}}{2}a}\int_{d=s}^{\frac{\sqrt{3}}{2}a}(d-s)f_{S_{1}}(s)f_{D_{1}}(d) \mathrm{d}d\mathrm{d}s & s,d\in \vartheta_{1} \\
		2\int_{s=0}^{\frac{\sqrt{3}}{2}a}\int_{d=\frac{\sqrt{3}}{2}a}^{\sqrt{3}a}(d-s)f_{S_{1}}(s)f_{D_{2}}(d) \mathrm{d}d\mathrm{d}s & s\in \vartheta_{1},d\in \vartheta_{2} \\
		2\int_{s=\frac{\sqrt{3}}{2}a}^{\sqrt{3}a}\int_{d=s}^{\sqrt{3}a}(d-s)f_{S_{2}}(s)f_{D_{2}}(d) \mathrm{d}d\mathrm{d}s & s,d\in \vartheta_{2} \\
	\end{cases}~,
\end{equation}
and we get $E(L) = \frac{41a}{45\sqrt{3}}$.
\subsection{APPENDIX-B}\label{APPENDIX-B}
\subsubsection{$E(L_{x})$ in $x$-axis}
Similarly, depending on the rang of $x$, we have:
\begin{itemize}
	\item [1)]  $x\in [0,\frac{1}{2}a]$, $\frac{1}{2}E_{1}(L_{x})=$
	\begin{equation}
		\begin{split}
			\begin{cases}
				\int_{s=0}^{x}\int_{d=s}^{x}(d-s)f_{S_{1}}(s)f_{D_{1}}(d)\mathrm{d}d\mathrm{d}s +\\ \int_{s=0}^{x}\int_{d=x}^{\frac{1}{2}a}(x-s)f_{S_{1}}(s)f_{D_{1}}(d)\mathrm{d}d\mathrm{d}s & s,d\in \theta_{1}\\
				\int_{s=0}^{x}\int_{d=\frac{1}{2}a}^{\frac{3}{2}a}(x-s)f_{S_{1}}(s)f_{D_{2}}(d)\mathrm{d}d\mathrm{d}s & d\in \theta_{2},s\in \theta_{1}\\
				\int_{s=0}^{x}\int_{d=\frac{3}{2}a}^{2a}(x-s)f_{S_{1}}(s)f_{D_{3}}(d)\mathrm{d}d\mathrm{d}s & d\in \theta_{3},s\in \theta_{1}\\
			\end{cases}~.
		\end{split}
	\end{equation}
	In the first case, $s,d\in [0,\frac{1}{2}a]$, as shown in Fig. \ref{lx-fig}, if $d=d_{1}$ (the case: $d\leq x$), $l_{x}(s,d) = d-s$ represents that the whole line segment from $s$ to $d$ is smaller than $x$. If $d=d_{2}$ (the case: $d> x$), $l_{x}(s,d) = x-s$ represents that the part line segment from $s$ to $x$ is smaller than $x$. There are only three piecewise cases in $E_{1}(L_{x})$, because $x\in [0,\frac{1}{2}a]$, for the other three cases, $s$ and $d$ are both greater than $x$, and thus $l_{x}(s,d) = 0$.
	\begin{figure}[htbp]
		\centering
		\includegraphics[width=0.5\textwidth]{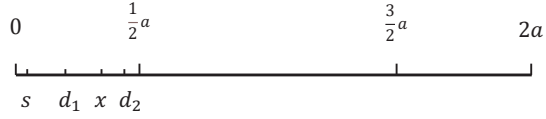}
		\caption{Illustration of the first case on line segment $[0,\frac{1}{2}a]$.}
		\label{lx-fig}
	\end{figure}
	\item [2)]  $x\in (\frac{1}{2}a,\frac{3}{2}a]$, $\frac{1}{2}E_{2}(L_{x})=$
	\begin{equation}
		\begin{split}
			\begin{cases}
				\int_{s=0}^{\frac{1}{2}a}\int_{d = s}^{\frac{1}{2}a}(d-s)f_{S_{1}}(s)f_{D_{1}}(d)\mathrm{d}d\mathrm{d}s &  s,d\in \theta_{1} \\
				\int_{s=0}^{\frac{1}{2}a}\int_{d = \frac{1}{2}a}^{x}(d-s)f_{S_{1}}(s)f_{D_{2}}(d)\mathrm{d}d\mathrm{d}s +\\ \int_{s=0}^{\frac{1}{2}a}\int_{d = x}^{\frac{3}{2}a}(x-s)f_{S_{1}}(s)f_{D_{2}}(d)\mathrm{d}d\mathrm{d}s &  s\in \theta_{1},d\in \theta_{2} \\
				\int_{s=\frac{1}{2}a}^{x}\int_{d = s}^{x}(d-s)f_{S_{2}}(s)f_{D_{2}}(d)\mathrm{d}d\mathrm{d}s +\\ \int_{s=\frac{1}{2}a}^{x}\int_{d = x}^{\frac{3}{2}a}(x-s)f_{S_{2}}(s)f_{D_{2}}(d)\mathrm{d}d\mathrm{d}s &  s,d\in \theta_{2} \\
				\int_{s=0}^{\frac{1}{2}a}\int_{d = \frac{3}{2}a}^{2a}(x-s)f_{S_{1}}(s)f_{D_{3}}(d)\mathrm{d}d\mathrm{d}s &  s\in \theta_{1},d\in \theta_{3} \\
				\int_{s=\frac{1}{2}a}^{x}\int_{d = \frac{3}{2}a}^{2a}(x-s)f_{S_{2}}(s)f_{D_{3}}(d)\mathrm{d}d\mathrm{d}s &  s\in \theta_{2},d\in \theta_{3} \\
			\end{cases}~.
		\end{split}
	\end{equation}
	
	\item [3)] $x\in (\frac{3}{2}a,2a]$, $\frac{1}{2}E_{3}(L_{x})=$
	\begin{equation}
		\begin{split}
			\begin{cases}
				\int_{s=0}^{\frac{1}{2}a}\int_{d=s}^{\frac{1}{2}a}(d-s)f_{S_{1}}(s)f_{D_{1}}(d)\mathrm{d}d\mathrm{d}s & s,d\in \theta_{1} \\
				\int_{s=0}^{\frac{1}{2}a}\int_{d=\frac{1}{2}a}^{\frac{3}{2}a}(d-s)f_{S_{1}}(s)f_{D_{2}}(d)\mathrm{d}d\mathrm{d}s & s\in \theta_{1},d\in \theta_{2} \\
				\int_{s=\frac{1}{2}a}^{\frac{3}{2}a}\int_{d=s}^{\frac{3}{2}a}(d-s)f_{S_{2}}(s)f_{D_{2}}(d)\mathrm{d}d\mathrm{d}s & s,d\in \theta_{2} \\
				\int_{s=0}^{\frac{1}{2}a}\int_{d=\frac{3}{2}a}^{x}(d-s)f_{S_{1}}(s)f_{D_{3}}(d)\mathrm{d}d\mathrm{d}s +\\ \int_{s=0}^{\frac{1}{2}a}\int_{d=x}^{2a}(x-s)f_{S_{1}}(s)f_{D_{3}}(d)\mathrm{d}d\mathrm{d}s & s\in \theta_{1},d\in \theta_{3} \\
				\int_{s=\frac{1}{2}a}^{\frac{3}{2}a}\int_{d=\frac{3}{2}a}^{x}(d-s)f_{S_{2}}(s)f_{D_{3}}(d)\mathrm{d}d\mathrm{d}s +\\ \int_{s=\frac{1}{2}a}^{\frac{3}{2}a}\int_{d=x}^{2a}(x-s)f_{S_{2}}(s)f_{D_{3}}(d)\mathrm{d}d\mathrm{d}s & s\in \theta_{2},d\in \theta_{3} \\
				\int_{s=\frac{3}{2}a}^{x}\int_{d=s}^{x}(d-s)f_{S_{3}}(s)f_{D_{3}}(d)\mathrm{d}d\mathrm{d}s +\\ \int_{s=\frac{3}{2}a}^{x}\int_{d=x}^{2a}(x-s)f_{S_{3}}(s)f_{D_{3}}(d)\mathrm{d}d\mathrm{d}s & s,d\in \theta_{3} \\
			\end{cases}~.
		\end{split}
	\end{equation}
\end{itemize}

\subsubsection{$E(L_{y})$ in $y$-axis}
Similarly, depending of the rang of $y$, we have:
\begin{itemize}
	\item [1)] $y\in [0,\frac{\sqrt{3}}{2}a]$, $\frac{1}{2}E_{1}(L_{y})=$
	\begin{equation}
		\begin{split}
			\begin{cases}
				\int_{s=0}^{y}\int_{d=s}^{y}(d-s)f_{S_{1}}(s)f_{D_{1}}(d)\mathrm{d}d\mathrm{d}s+\\
				\int_{s=0}^{y}\int_{d=y}^{\frac{\sqrt{3}}{2}a}(y-s)f_{S_{1}}(s)f_{D_{1}}(d)\mathrm{d}d\mathrm{d}s & s,d\in \vartheta_{1} \\
				\int_{s=0}^{y}\int_{d=\frac{\sqrt{3}}{2}a}^{\sqrt{3}a}(y-s)f_{S_{1}}(s)f_{D_{2}}(d)\mathrm{d}d\mathrm{d}s & s\in \vartheta_{1},d\in \vartheta_{2} \\
			\end{cases}~.
		\end{split}
	\end{equation}
	\item [2)] $y\in (\frac{\sqrt{3}}{2}a,\sqrt{3}a]$, $\frac{1}{2}E_{2}(L_{y})=$
	\begin{equation}
		\begin{split}
			\begin{cases}
				\int_{s=0}^{\frac{\sqrt{3}}{2}a}\int_{d=s}^{\frac{\sqrt{3}}{2}a}(d-s)f_{S_{1}}(s)f_{D_{1}}(d)\mathrm{d}d\mathrm{d}s & s,d\in \vartheta_{1} \\
				\int_{s=0}^{\frac{\sqrt{3}}{2}a}\int_{d=\frac{\sqrt{3}}{2}a}^{y}(d-s)f_{S_{1}}(s)f_{D_{2}}(d)\mathrm{d}d\mathrm{d}s +\\\int_{s=0}^{\frac{\sqrt{3}}{2}a}\int_{d=y}^{\sqrt{3}a}(y-s)f_{S_{1}}(s)f_{D_{2}}(d)\mathrm{d}d\mathrm{d}s & s\in \vartheta_{1},d\in \vartheta_{2} \\
				\int_{s=\frac{\sqrt{3}}{2}a}^{y}\int_{d=s}^{y}(d-s)f_{S_{2}}(s)f_{D_{2}}(d)\mathrm{d}d\mathrm{d}s +\\\int_{s=\frac{\sqrt{3}}{2}a}^{y}\int_{d=y}^{\sqrt{3}a}(y-s)f_{S_{2}}(s)f_{D_{2}}(d)\mathrm{d}d\mathrm{d}s & s,d\in \vartheta_{2} \\
			\end{cases}~.
		\end{split}
	\end{equation}
\end{itemize}

\bibliographystyle{abbrv}
\bibliography{TechnicalReport}

\end{document}